\documentclass[aip,apl,reprint]{revtex4-1}
\usepackage{amsmath,graphicx}
\draft

\begin{document}
	
\title{Versatile photon gateway based on controllable multiphoton blockade}
	
\author{Kui Hou}
\affiliation{MOE Key Laboratory of Advanced Micro-Structured Materials, School of Physics Science and Engineering,Tongji University, Shanghai, China 200092}
\affiliation{Department of Mathematics and Physics, Anhui JianZhu University, Hefei, China 230601}
	
\author{Jizi Lin}
\affiliation{MOE Key Laboratory of Advanced Micro-Structured Materials, School of Physics Science and Engineering,Tongji University, Shanghai, China 200092}
\affiliation{Department of physics,Zhangjiagang campus, Jiangsu university of science and technology, Zhangjiagang, China 215600}
	
\author{Chengjie Zhu}
\email[Corresponding author:]{cjzhu@tongji.edu.cn}
\affiliation{MOE Key Laboratory of Advanced Micro-Structured Materials, School of Physics Science and Engineering,Tongji University, Shanghai, China 200092}
	
\author{Yaping Yang}
\email[Corresponding author:]{yang\_yaping@tongji.edu.cn}
\affiliation{MOE Key Laboratory of Advanced Micro-Structured Materials, School of Physics Science and Engineering,Tongji University, Shanghai, China 200092}
	
\begin{abstract}
Manipulating photons is an essential technique in quantum communication and computation. Combining the Raman electromagnetically induced transparency technology, we show that the photon blockade behavior can be actively controlled by using an external control field in a two atoms cavity-QED system. As a result, a versatile photon gateway can be achieved in this system, which changes the cavity photons from classical to quantum property and allows one photon, two photon and classical field leaking from the cavity. The proposal presented here has many potential applications for quantum information processing and can also be realized in circuit QED system.
\end{abstract}

\maketitle

\noindent Photons are natural carriers of photonic quantum information processing which allows to
deliver sensitive data for security communication~\cite{gisin2007quantum}. It is shown that the transfer rate of quantum communications scales exponentially with the photon flux over long distances~\cite{eisaman2011invited}. Using the quantum network technology~\cite{kimble2008hj}, it is possible to realize quantum computation with photons~\cite{albash2018adiabatic}, which can solve certain problems much more efficiently than any classical computers. One of the key factors to realize quantum information communication and computation is the achievement of few-photon sources, especially the single photon as resources~\cite{lounis2005single}. 

A typical method to achieve single photon emission is based on the two-photon blockade phenomenon~\cite{PhysRevLett.79.1467}, which absorbs the first photon, but blocks the absorption of the next photon due to the presence of the photon-number-dependent frequency shift. This phenomenon was first experimentally observed in an optical cavity strongly coupled to a single atom~\cite{birnbaum2005photon}. Using this method, single-photon emission has been theoretically proposed and experimentally demonstrated in many systems, including semiconductor quantum dots~\cite{michler2000quantum}, solid state~\cite{ding2016on-demand}, light-matter interaction~\cite{wu2017bright}, spectral multiplexing~\cite{puigibert2017heralded}, the cavity QED system~\cite{hennessy2007quantum},
the circuit QED systems~\cite{PhysRevLett.106.243601},
optomechanical systems~\cite{PhysRevLett.107.063601} and so on. Moreover, Rempe et al. showed that atom-cavity system transforms a random stream of input photons into a correlated stream of output photons~\cite{kubanek2008two} and two-photon blockade can be optically controlled by using the electromagnetically induced transparency (EIT) technology~\cite{PhysRevLett.111.113602}, which opens up the possibility to realize the two-photon gateway.

Although the two-photon blockade phenomenon has been extensively investigated, there are few works on the multiphoton blockade, especially the three-photon blockade, which is still a challenge in both theory and experiments. A direct method to realize three-photon blockade is by increasing the pump field intensity so that two-photon excitations can be strong enough to be observed. However, the power broadening caused by the pump field eliminate the energy difference of each doublet so that multiphoton excitations take place simultaneously. As a consequence, the condition of the three-photon blockade is very strict and difficult to be realized in experiments~\cite{PhysRevLett.118.133604}. Recently, we propose a novel system to achieve three-photon blockade in a single mode cavity QED system containing two identical two-level atoms with different coupling strengths~\cite{PhysRevA.95.063842}. When these two atoms radiate with out phase, one-photon excitations are forbidden so that the two-photon excitations are dominant. Therefore, three-photon blockade with reasonable photon number can be observed in our system. 

Combining the Raman-EIT technique with this two atoms cavity QED system~\cite{PhysRevLett.97.233001}, we show in this work that the multiphoton blockade behavior can be optically controlled by an external control field. We note that a similar scheme has already been experimentally studied where a single atom is trapped in the cavity and the control field is resonant with the cavity to form an EIT configuration~\cite{PhysRevLett.111.113602}. However, we find many interesting features of the photon blockade associated with the collective radiations in this Raman EIT configuration that a single atom and the EIT scheme do not possess. Choosing a specific pump field frequency, we show that the property of the photons leaking from the cavity can be changed from bunching to anti-bunching behavior by tuning the control field Rabi frequency. We further show that, in the case of in-phase radiations, the two-photon blockade can be changed to three-photon blockade by just varying the control field intensity. Therefore, it is possible to actively control the photon emissions in this system. Based on these characteristics, we show that a versatile photon gateway can be realized in our system, which can be achieved in experiments with current experimental conditions.  

\begin{figure}[t]
	\centering
	\includegraphics[width=8.0cm]{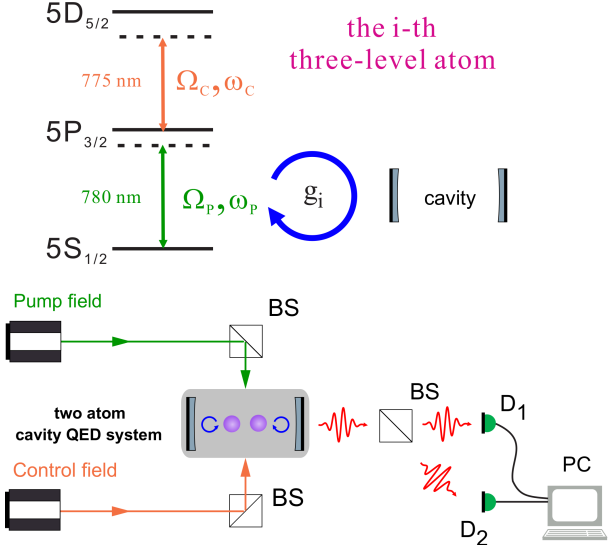}
	\caption{The schematic of two identical three-level atoms trapped in a single mode cavity with different coupling strength $g_i$ ($i=1,2$). A pump field $\Omega_P$ with angular frequency $\omega_P$ couples the $|g\rangle\leftrightarrow|m\rangle$ transition, and a control field $\Omega_{\rm C}$ with angular frequency $\omega_c$ couples the $|m\rangle\leftrightarrow|e\rangle$, forming a ladder-type configuration. $\Delta_p$ and $\gamma_m$ ($\Delta_c$ and $\gamma_e$) are the detuning of pump (control) field and decay rate of the state $|m\rangle$ ($|e\rangle$), respectively. $\kappa$ stands for the cavity decay rate.}
	\label{fig:model}
\end{figure}

We consider an atom-cavity QED system where two identical three-level atoms are trapped in a single mode cavity with different coupling strengths. The corresponding energy levels are labeled as $|g\rangle\equiv|{\rm 5S_{1/2}}\rangle$, $|m\rangle\equiv|{\rm 5P_{3/2}}\rangle$ and $|e\rangle\equiv|{\rm 5D_{5/2}}\rangle$, respectively. A pump field with Rabi frequency $\Omega_P$ couples the $|g\rangle\leftrightarrow|m\rangle$ transition, and another control field with Rabi frequency $\Omega_c$ couples the $|m\rangle\leftrightarrow|e\rangle$ transition (see Fig.~\ref{fig:model}). For simplicity, we assum the cavity mode frequency $\omega_{\rm cav}$ is the same as the transition frequency between state $|g\rangle$ and $|m\rangle$, i.e., $\omega_{\rm cav}=\omega_{m}-\omega_{g}$ with $\hbar\omega_{\alpha}\ (\alpha=g,m,e)$ being the energy of state $|\alpha\rangle$.

Under the electric dipole and rotating wave approximation, the Hamiltonian of the system is written as
\begin{eqnarray}\label{eq:H}
H&=&\hbar[\sum_{j=1}^2(\Delta_mS^j_{mm}
+\Delta_eS^j_{ee})+\Delta_{\rm cav} a^\dag a \nonumber\\
&+&\sum_{j=1}^2(g_jaS^j_{mg}+\Omega_PS^j_{mg}+\Omega_CS^j_{em}+{\rm H.C})]
\end{eqnarray}
where the detuning is defined as $\Delta_{\rm cav}=\omega_{\rm cav}-\omega_{\rm p}$, $\Delta_m=\Delta_{\rm p}$ and $\Delta_e=\Delta_{\rm p}+\Delta_c$ with  $\Delta_{\rm p}=\omega_m-\omega_g-\omega_{\rm p}$ and $\Delta_c=\omega_e-\omega_m-\omega_c$ being the detuning of pump and control field, respectively. Here, $S^j_{\alpha\beta}=|\alpha\rangle_j\langle\beta|\ (\alpha,\beta=\{g,m,e\})$ is the j-th atomic operator and $a$ ($a^\dag$) is the photon annihilation (creation) operator. The position-dependent coupling strength of the $j$-th atom $g_j=g\cos(2\pi z_j/\lambda_c)\ (j=1,2)$, where $z_j$ is the position of the $j$-th atom and $\lambda_{\rm cav}=\omega_{\rm cav}/c_0$ is the wavelength of the cavity mode ($c_0$ is the light speed in vacuum)~\cite{PhysRevA.76.053829}.

\begin{figure}[t]
	\includegraphics[width=8.0cm]{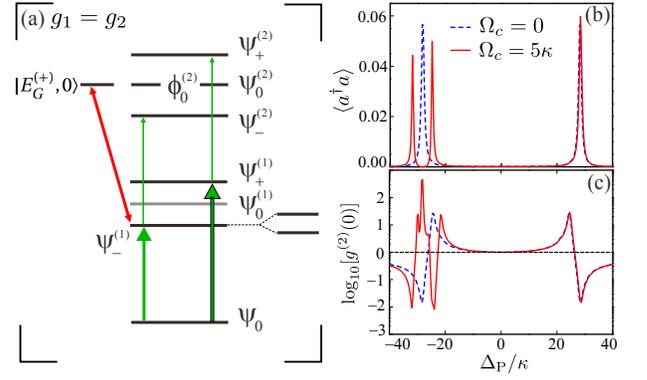}
	\caption{(a) Diagram of the dressed states structure for $g_1=g_2$. Here, only main two-photon transitions are shown in panel (a). Panels (b) and (c) show the mean photon number $\langle a^\dag a\rangle$  and the equal-time second-order field correlation function in log unit $\log_{10}[g^{(2)}(0)]$ as a function of the pump field detuning, respectively. The control field is chosen as $\Omega_C=0$ (blue dashed curves) and $\Omega_C=5\kappa$ (red solid curves), respectively. Other system parameters are given in the text. The black dash-dotted line indicates $g^{(2)}{(0)}=1$. Other parameters are  $g=20\kappa$, $\Omega_p=0.2\kappa$, $\Delta_c=\sqrt{2}g$, $\gamma_{m}=\kappa$ and $\gamma_{e}=\kappa/100$, respectively.
}
	\label{fig:enery split1}
\end{figure}
\begin{figure*}[htb]
	\includegraphics[width=16cm]{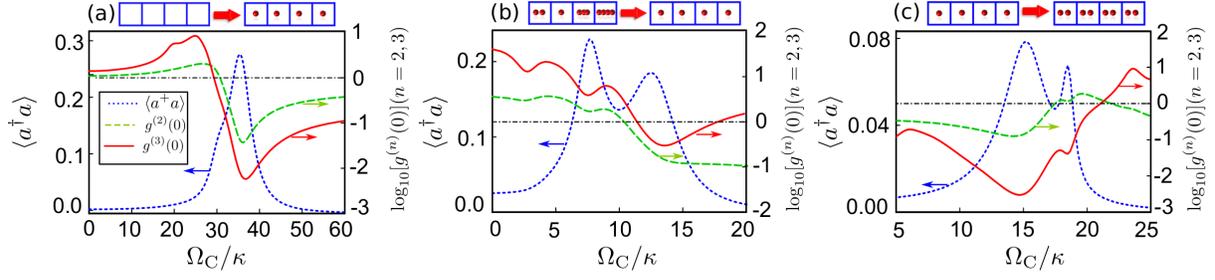}
	\caption{Plots of mean photon number (blue dotted curves), field correlation functions $g^{(2)}(0)$ (green dashed curves) and $g^{(3)}(0)$ (red solid curve) as a function of the control field Rabi frequency $\Omega_C$ with $g_1=g_2$. Here, we choose $\Omega_P=1.5\kappa$ and the pump field detuning is given by $\Delta_p=-10\kappa$ [panels (a)], $\Delta_p=-20\kappa$ [panels (b)] and $\Delta_p=-40\kappa$ [panels (c)], respectively. The black dash-dotted line indicates $g^{(2)}(0)=g^{(3)}(0)=1$.}
	\label{fig:gateway1}
\end{figure*}
In general, the dynamical evolution of the system can be described by using the master equation, i.e.,
\begin{align}\label{eq:master}
&\frac{\partial\rho}{\partial t}=-\frac{i}{\hbar}[H,\rho]
+\kappa\mathcal{L}_\kappa\rho
+\sum_{j=1}^2\left[\gamma_{e}\mathcal{L}_j\rho+\gamma_{m}\mathcal{L}_j^{'}\rho\right],
\end{align}
where
\begin{subequations}\label{eq:atom decay}
\begin{eqnarray}
\mathcal{L}_\kappa\rho&=&a\rho a^\dag-(a^\dag a\rho+\rho a^\dag a)/2,\\
\mathcal{L}_j\rho&=&S^j_{em}\rho S^j_{me}-(S^j_{em}S^j_{me}\rho+\rho S^j_{em}S^j_{me})/2,\\
\mathcal{L}_j^{'}\rho&=&S^j_{mg}\rho S^j_{gm}-(S^j_{mg}S^j_{gm}\rho+\rho S^j_{mg}S^j_{gm})/2,
\end{eqnarray}
\end{subequations}
are the superoperators that describe the quantum noise due to the cavity decay and the spontaneous emissions of the j-th atom. $\kappa$ is the cavity decay rate and $\gamma_m$ ($\gamma_e$) is the spontaneous decay rate of atomic state $|m\rangle$ ($|e\rangle$). Solving Eq.~(2) numerically, we can obtain many interesting quantum features of the cavity field, which will be demonstrated in the following. It is noted that the quantum properties of the cavity field are strongly dependent to the positions of two atoms~\cite{pleinert2017hyperradiance}. 

First, we consider that two atoms have the same coupling strengths, i.e., $g_1=g_2=g$. In this case, it is difficult to see the physical process by using the dressed state picture, which can be obtained by solving the Hamiltonian of the whole system. To show the physical mechanism more clearly, we decompose the system into two components. One is a subsystem consisting of the cavity and two two-level atoms, and the other is a subsystem consisting of the control field and the collective state associated with the excited state $|e\rangle$, i.e., $|E_G^\pm\rangle=(|eg\rangle\pm|ge\rangle)/\sqrt{2}$, $|E_M^\pm\rangle=(|em\rangle\pm|me\rangle)/\sqrt{2}$ and $|ee\rangle$. The first subsystem has been studied in many literatures~\cite{PhysRevA.95.063842}, and the corresponding dressed state energies and eigenstates can be obtained easily by introducing the collective states $|gg\rangle$,  $|M_g^\pm\rangle=(|mg\rangle\pm|gm\rangle)/\sqrt{2}$ and $|mm\rangle$ as basis. In one photon space, we obtain a set of eigenstates $\Psi^{(1)}_{\pm}=(|M_g^+,0\rangle\pm|gg,1\rangle)/\sqrt{2}$ with eigenvalues $\lambda^{(1)}_{\pm}=\hbar\omega_c\pm\sqrt{2}g\hbar$. Likewise, in two photon space, we can also obtain a set of eigenstates $\Psi^{(2)}_0=(-|gg,2\rangle+\sqrt{2}|ee,0\rangle)/\sqrt{3}$ with eigenvalue $\lambda^{(2)}_0=2\hbar\omega_c$, and $\Psi^{(2)}_\pm=(2\sqrt{3}|gg,2\rangle+\sqrt{6}|ee,0\rangle\pm3\sqrt{2}|M_g^+,1\rangle)/6$ with eigenvalues  $\lambda^{(2)}_\pm=2\hbar\omega_c\pm\sqrt{6}g\hbar$ [see Fig.~\ref{fig:enery split1}(a)].

It is clear to see that there exist two one-photon excitation pathways when the control field is turned off. Correspondingly, there are two peaks in the cavity excitation spectrum can be observed with two-photon blockade behavior, i.e., $g^{(2)}(0)\equiv\langle a^\dag a^\dag aa\rangle/\langle a^+a\rangle^2<1$ [see blue curves in Fig.~2(b) and 2(c)]. If one turn on the control field and set its frequency resonant to the $\Psi_-^{(1)}\leftrightarrow|E_G^{(+)}\rangle$ transition (e.g., $\Delta_c=-\sqrt{2}g$), the state $\Psi_-^{(1)}$ will be split into two states due to the coupling of the control field. We note that the energy of the state $\Psi_+^{(1)}$ changes slightly because the control field couples the $\Psi_-^{(1)}\leftrightarrow|E_G^{(+)}\rangle$ transition nonresonantly. As a result, one can observe three peaks with two-photon blockade [i.e., $g^{(2)}(0)<1$] behavior in the cavity excitation spectrum and the distance between two left side peaks is proportional to the control field Rabi frequency. This method provide a possibility to active control the quantum property of the cavity field and the statistic behavior of the photons emitted from the cavity, resulting in one-photon gateway operations~\cite{PhysRevA.84.033829,javadi2015single}.

\begin{figure}[b]
	\includegraphics[width=8.5cm]{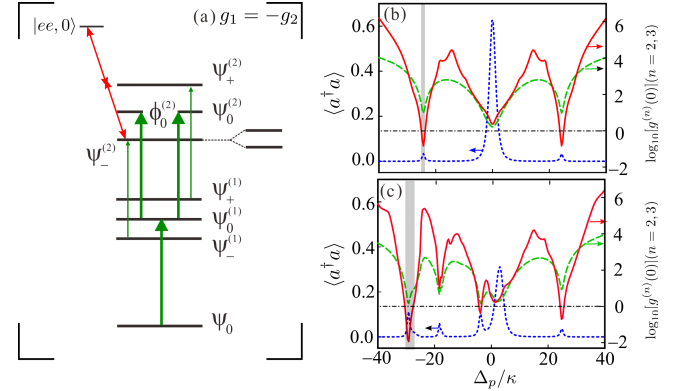}
	\caption{(a) Diagram of the dressed states structure for $g_1=-g_2$. Here, only main two-photon transitions are shown in panel (a). The mean photon number (blue dotted curve), equal-time field correlation functions $g^{(2)}(0)$ (green dashed curves) and $g^{(3)}(0)$ (red solid curves) are shown in panels (b) and (c). The control field is chosen as $\Omega_C=0$ (b) and $\Omega_C=8\kappa$ (c), respectively. Here, we choose $\Omega_{\rm P}=2\kappa$, $g=20\kappa$, $\Delta_c=\sqrt{6}g/2$ and other system parameters are the same as those used in Fig. 2. The gray area indicate the regime of three-photon blockade, and the black dash-dotted line indicates $g^{(2)}(0)=g^{(3)}(0)=1$.}
	\label{fig:enery split2}
\end{figure}
\begin{figure*}[t]
	\includegraphics[width=16cm]{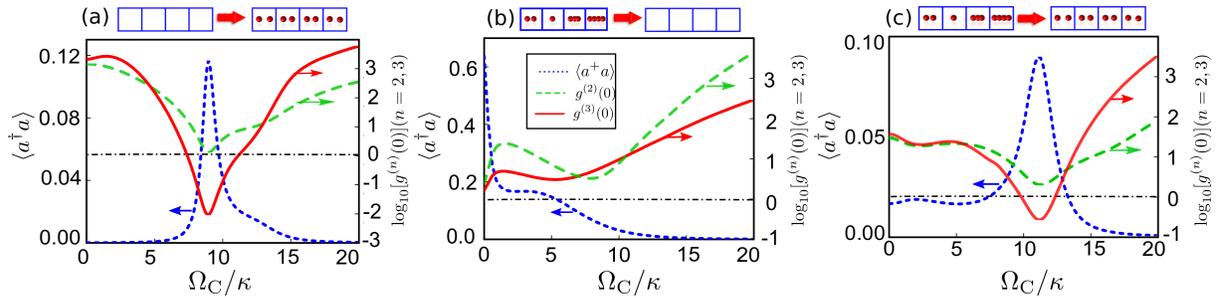}
	\caption{Plots of mean photon number (blue dotted curves), field correlation function $g^{(2)}(0)$ (green dashed curves) and $g^{(3)}(0)$ (red solid curve) versus the control field Rabi frequency $\Omega_C$ with $g_1=-g_2$. In panels (a), (b) and (c), we choose $\Omega_{\rm p}=2\kappa$ and the the pump field detuning are given by $\Delta_{\rm p}=-30\kappa$, $\Delta_{\rm p}=0$ and $\Delta_{\rm p}=-5\kappa$, respectively. The black dash-dotted line indicates $g^{(2)}(0)=g^{(3)}(0)=1$.}
	\label{fig:gateway2}
\end{figure*}
For example, taking the pump field detuning $\Delta_{\rm p}=-10\kappa$, few photons can leak from the cavity in the case of weak control field since the pump field is far off-resonant with the $\Psi_0\leftrightarrow\Psi^{(1)}_-$ transition. Increasing the control field Rabi frequency, the state $\Psi^{(1)}_-$ is split into two states, and one of them becomes resonant to the pump field, resulting in one-photon emission due to the two-photon blockade [see Fig.~3(a)]. Here, the system parameters are given by $g=20\kappa$, $\Omega_{\rm p}=1.5\kappa$, $\gamma_{m}=\kappa$ and $\gamma_{e}=\kappa/100$, respectively. However, if one take $\Delta_{\rm p}=-20\kappa$ (close to the one-photon excitation frequency), the property of the cavity photon changes from bunching to antibunching behavior as the control field increases [see Fig.~3(b)]. More interesting, at the frequencies of two-photon excitations, the three-photon blockade [i.e., $g^{(2)}(0)>1$ and $g^{(3)}(0)\equiv\langle a^\dag a^\dag a^\dag aaa\rangle/\langle a^\dag a\rangle^3<1$] takes place. As shown in Fig.~3(c), one-photon emission can be changed to two-photon emission by just increasing the control field intensity when the pump field detuning is chosen as $\Delta_p=-40\kappa$.

Next, we consider the case of $g_1=-g_2$ where the one-photon excitations are forbidden and two-photon excitations are dominant [see Fig. 4(a)]. Therefore, three-photon blockade can be observed at the frequencies of two-photon excitations as shown in Fig. 4(b). Taking the control field detuning $\Delta_c=-\sqrt{6}g/2$, the two-photon excitation state $\Psi^{(2)}_-$ is split into a doublet since it is coupled to the state $|ee,0\rangle$ by the control field via two-photon process [see panel (a)]. Consequently, the three-photon blockade can be significantly improved and the frequency regime to realize three-photon blockade can be broadened as shown in Fig.~4(c).

Furthermore, multiphoton gateway operations can also be achieved under the condition of $g_1=-g_2$. When the pump field is nonresonant (e.g., $\Delta_{\rm p}=-30\kappa$), there is few photon can be measured. Increasing the control field Rabi frequency, two-photon excitation state $\Psi^{(2)}_-$ is split into a doublet and the pump field is resonant to the two-photon excitation state so that antibunching photons with three-photon blockade behavior can be measured [see Fig. 5(a)]. As shown in Fig. 5(b), taking the pump field detuning $\Delta_{\rm p}=0$, cavity photons exhibit bunching behavior since the multiphoton excitations are allowed if the control field is very weak. As the control field Rabi frequency increases, the pump field becomes nonresonant to the cavity because of the energy shift induced by the control field. Therefore, few photon can be detected. For a small pump field detuning (e.g., $\Delta_{\rm p}=-5\kappa$), the property of the cavity field can be changed from classic (bunching) to three-photon blockade behavior by increasing the control field intensity [see panel (c)]. 

In summary, we have studied the two atoms cavity QED system, where two atoms are directly driven by a pump field and a control field, forming a Raman EIT configuration. We show that the photon blockade can be optically controlled by using the control field to couple the one- and two-photon excitation states. Base on this method, we show that a versatile photon gateway operations can be achieved in our system, allowing one photon, two photons and classical field leaking from the cavity. We also show that the property of the cavity photon can be changed from bunching to antibunching by using the gateway operations. This versatile photon gateway holds many potential applications in quantum communication and computation, and it not only can be achieved in atom-cavity QED system, but also can be realized in artificial atom system, for example, the circuit-QED system, quantum-dot-cavity QED system and so on.

{\bf Acknowledgments} This work was supported by the National Key Basic Research Special Foundation (2016YFA0302800); the Shanghai Science and Technology Committee (18JC1410900); the National Nature Science Foundation (11774262); the Natural Science Foundation of Anhui Province (1608085QA23).

\bibliography{ref}

\end{document}